\documentclass{article}

\usepackage{graphicx}
\usepackage{float}
\usepackage{amsmath,amssymb,amstext}

\begin{document}

\title{Characterization of charge collection in CdTe and CZT using the transient current technique}

\author{J.~Fink \footnote{Physikalisches Institut, Nussallee 12,
              D-53115 Bonn, Germany, Tel.: +49\,228\,73-3533, Fax:
               -3220, email: fink@physik.uni-bonn.de}, H.~Kr{\"u}ger, P.~Lodomez and N. Wermes}

\date{21. November 2005}

\maketitle

\begin{abstract}
The charge collection properties in different particle sensor materials with respect to the shape of the generated signals, the electric field within the detector, the charge carrier mobility and the carrier lifetime are studied with the transient current technique (TCT). Using the well-known properties of Si as a reference, the focus is laid on Cadmium-Telluride (CdTe) and Cadmium-Zinc-Telluride (CZT), which are currently considered as promising candidates for the efficient detection of X-rays.\\
All measurements are based on a transient-current technique (TCT) setup, which allows the recording of current pulses generated by an $^{241}$Am $\alpha$-source. These signals will be interpreted with respect to the build-up of space-charges inside the detector material and the subsequent deformation of the electric field. Additionally the influence of different electrode materials (i.e. ohmic or Schottky contacts) on the current pulse shapes will be treated in the case of CdTe. Finally, the effects of polarization, i.e. the time-dependent degradation of the detector signals due to the accumulation of fixed charges within the sensor, are presented.
\end{abstract}

\section{Introduction}

While Si is the standard sensor material for micro strip or pixel detectors for charged particle detection in high energy physics, its application to X-ray imaging is limited due to its low atomic number. Nevertheless, the concept of a directly converting sensor material in combination with a pixellated ASIC readout chip has caused a large interest in alternative semiconductor materials like CdTe and CZT. In terms of stopping power these two materials profit from their high atomic numbers (Z$_{Cd}$ = 48, Z$_{Zn}$ = 30 and Z$_{Te}$ = 50), but until recently their application for radiation detection has been limited due to the reduced material quality with respect to the collection of charges generated inside the material. With high quality CdTe and CZT now being commercially available, detailed studies of the charge carrier transport within the sensor material have been carried out in this paper.\\
Generally, signal generation by ionizing radiation in a semiconductor detector is based on the creation of electron-hole pairs. The subsequent detection of the deposited charge is realized through the application of a potential difference between the metal contacts, which causes the generated charge carriers to drift towards the oppositely charged electrode. In the case of a single channel, parallel plate detector this means, that the charge carrier movement causes a measurable current signal immediately after the generation and separation of the electron-hole pairs. Practically all high energy particle experiments integrate this current via a charge sensitive amplifier, yielding an output voltage proportional to the created charge. This is where the transient-current technique differs from the common approach. In a TCT setup a fast readout chain is used to directly amplify the current signal, as the charge carriers travel through the detector. The advantage of such a time-resolved current measurement over the common charge-sensitive approach is the ability to directly map the charge carrier movement within the detector material without any integration of the signal current.\\
The discussion of the experimental results starts with a short introduction of the TCT-setup, followed by the presentation of the measured current pulses in section \ref{sec:PulseShape}. Sections \ref{sec:Mob}, \ref{sec:ElField} and \ref{sec:Pol} deal with the analysis of the current signals.

\section{The experimental setup}
\begin{figure}[H]
\begin{center}
\includegraphics[width=0.75\textwidth]{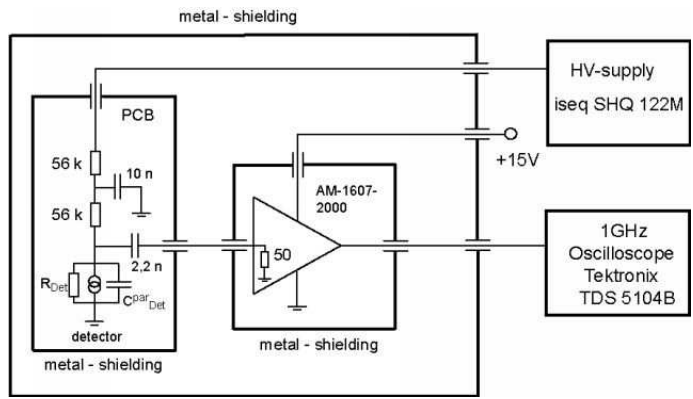}
\caption{Schematic view of the TCT-setup. The detector is replaced by its equivalent circuit diagram consisting of the detector resistance R$_{Det}$, the parasitic capacitance C$^{par}_{Det}$ ($\simeq$~1~pF) and a current source.}
\label{fig:Fig1.eps}
\end{center}
\end{figure}
Figure \ref{fig:Fig1.eps} shows a schematic view of the experiment. A small, resistance-matched and shielded PCB provides the biasing network and the socket for the detector crystals. The signals coming from the sensor are AC-coupled to a commercial voltage amplifier (Miteq AM-1607-2000) with a gain of 41~dB. An ionizing particle creates a current signal within the detector. This current pulse is converted into a voltage pulse through the input impedance (50~$\Omega$) of the voltage amplifier, giving an overall trans\-impedance gain of 5840 mV$/$$\mu$A. The systems voltage noise is (3.5~$\pm$~0.5) mV$_{RMS}$ at 2~GHz bandwidth. The amplified voltage pulses are stored in a broadband digital oscilloscope (Tektronix TDS 5104B 1~GHz). For further improvement of the noise characteristics, especially the quenching of electro-magnetic pickup a second shielding box is placed around the PCB and the amplifier.\\   
All of the following measurements use 5.5~MeV $\alpha$-particles from an $^{241}$Am-source in order to create electron-hole pairs within the detector. The main reason for this is the short penetration depth of $\alpha$-particles in matter (approx. 10-20 $\mu$m in CdTe), which guarantees a signal generation close to the irradiated electrode. This in turn enables the observation of purely electron or purely hole induced signals. In both cases one type of charge carriers traverses the whole detector volume and thus generates the signal, whereas the oppositely charged type does not contribute to the signal as these carriers almost instantly reach the collecting electrode. Additionally, the limited range of $\alpha$-particles allows the averaging over many current pulses, because the starting conditions for each charge carrier migration are the same for all events.\\
Within the short range of $\alpha$-particles also lies their major disadvantage, as the energy loss in the air and inside the source material itself cannot be neglected. Measurements with a conventional charge sensitive setup yield an average energy loss of 1.5~MeV~$\pm$~0.05~MeV for a detector-source distance of 10~mm air and a remaining $\alpha$-particle energy of 3.9~MeV~$\pm$~0.05~MeV (corresponding to 141~fC~$\pm$~1~fC in CdTe). By extrapolating the detector source distance to zero it is possible to determine the energy at which the $\alpha$-particles leave the $^{241}$Am-source to about 4.7~MeV.\\

\section{Current pulses}
\label{sec:PulseShape}
Current pulses in any kind of particle detector, whose operation is based on the induction of mirror charges on a certain number of electrodes, can be described by the Ramo-Shockley theorem \cite{Ramo39:Lit,Shockley38:Lit}.
\begin{equation}
\label{eqn:Eqn1}
i(t) = Q_{e-}(t) \cdot E_{W} \cdot v_{drift}^{e-}(x(t)) + Q_{h+}(t) \cdot E_{W} \cdot v_{drift}^{h+}(x(t))
\end{equation}
With i being the signal current, Q the electron or hole charge, E$_W$ the weighting field and v$_{drift}$ the drift velocity. E$_W$ only describes the coupling of the charge carrier movement to the readout electrode and is not to be confused with the electric field E(x(t)), which determines the trajectory of the particles inside the detector. In the case of a single channel detector with parallel electrodes at a distance D, the expression for the weighting field is reduced to the simple term 1/D. Solving the equation of motion for the created charge carriers under the assumption of a linear electric field distribution (caused by a constant space-charge) yields an exponential current signal \cite{Lutz99:Lit}.\\
Apart from a constant space-charge density inside the sensor material, charge carrier trapping can also influence the pulse shape. Equation (\ref{eqn:Eqn1}) states that the current amplitude i(t) is proportional to the deposited charge Q(t). Hence an exponential decay of the charge inside the detector again yields an exponential decay of the current amplitude \cite{Zanio68:Lit}.

\subsection{Silicon}

\begin{figure}[h]
\begin{center}
\includegraphics[width=0.75\textwidth]{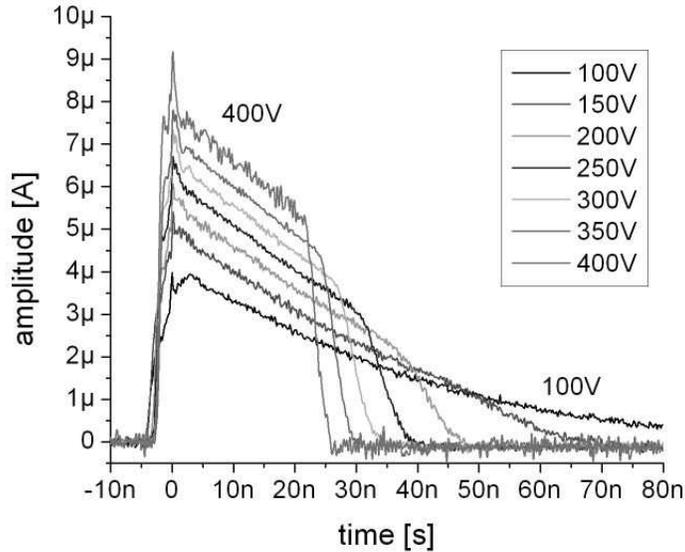}
\caption{Electron signals in a 1~mm Si p-n diode.$\alpha$-particles impinging on the cathode.}
\label{fig:Fig2.eps}
\end{center}
\end{figure}

The properties of Si p-n diodes have previously been studied in detail \cite{Lutz99:Lit,Krs04:Lit,Jacoboni76:Lit}. In this work Si p-n diodes were used as reference devices for the studies on CdTe and CZT. Figure \ref{fig:Fig2.eps} shows electron induced current pulses in a 1~mm thick Si diode, irradiated from the cathode (p+) side. The pulse durations range from 140~ns at 100~V down to 25~ns at 400~V, with maximum currents between 4~$\mu$A and 8~$\mu$A. 
For high voltages the rising edge of the signal is dominated by the signal electronics and the separation of the charge carriers (see section \ref{subsec:E-CdTeO}). This initial rising edge is followed by an exponential decay, caused by the negative space-charge inside the weakly doped n-type material. The exponential decay ends upon the arrival of the first electrons at the anode, which can be seen as a more or less prominent bend in the current signals (indicated by the arrow in Fig. \ref{fig:Fig2.eps}). The subsequent final drop of the amplitude is governed by the longitudinal diffusion of the charge carriers, which has taken place during their movement through the detector. For low voltages the charge carrier cloud can reach larger dimensions and, as a consequence, the signal falls off slowly after the first carriers arrive at the electrode. The spike at t = 0 is a trigger artifact. The full depletion voltage of these diodes was determined by capacity measurements (C$_{FD}$ = 12~pF~$\pm$~1~pF) and by fits to the pulse shape and lies at 96~V~$\pm$~5~V.
\begin{figure}
\begin{center}
\includegraphics[width=0.75\textwidth]{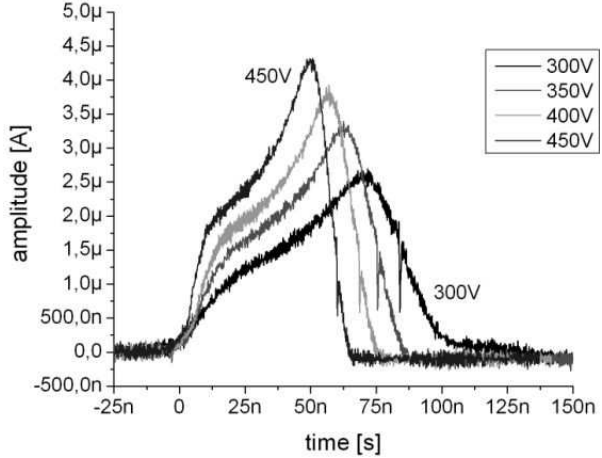}
\caption{Hole signals in a 1~mm Si p-n diode.$\alpha$-particles impinging on the anode.} 
\label{fig:Fig3.eps}
\end{center}
\end{figure}       
For voltages sufficiently above the full depletion bias (V$_{bias}>$~300~V) hole signals are observed. Figure \ref{fig:Fig3.eps} shows signals that were recorded with the anode (n-side) being irradiated by the $\alpha$-particles. The significant delay at the beginning of the signal is caused by the low electric field in the anode region.

\subsection{Cadmium-Telluride}

\begin{figure}[H]
\begin{center}
\includegraphics[width=0.75\textwidth]{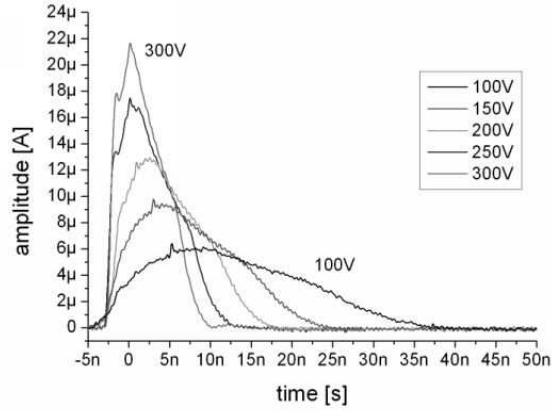}
\caption{Electron signals in \mbox{CdTe-O} with cathode irradiation (D~=~500~$\mu$m). }
\label{fig:Fig4.eps}
\end{center}
\end{figure}

\begin{figure}[H]
\begin{center}
\includegraphics[width=0.75\textwidth]{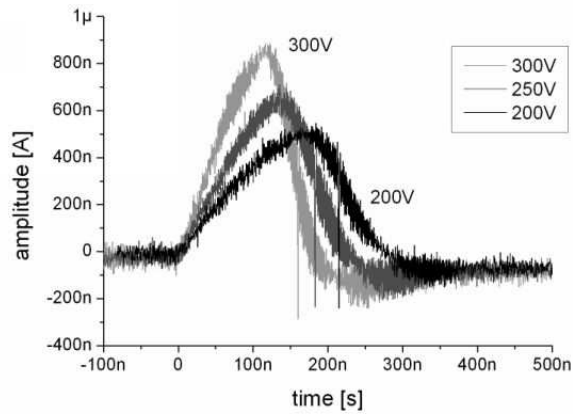}
\caption{Hole signals in \mbox{CdTe-O} with anode irradiation (D~=~500~$\mu$m).}
\label{fig:Fig5.eps}
\end{center}
\end{figure}

A total of four CdTe detector crystals with two different electrode configurations were analyzed with the presented TCT-setup. The first pair, named \mbox{CdTe-O}, has Platinum (Pt) electrodes on both sides, providing an ohmic contact behavior. The second set of crystals, here labelled as \mbox{CdTe-S}, has an Indium (In) electrode on the backside and a regular Pt contact on the topside. All available CdTe and CZT samples are glued to a ceramic holder, thus allowing only the irradiation of the top Pt electrode. Accordingly, hole signals cannot be observed with the \mbox{CdTe-S} sensors, as these detectors need a reverse bias for operation.
Figures \ref{fig:Fig4.eps} and \ref{fig:Fig5.eps} show electron signals in \mbox{CdTe-O} and \mbox{CdTe-S} for different bias voltages. It is evident that while the maximum amplitudes for identical bias settings are of comparable height, the pulse shapes clearly depend on the type of contact electrodes. 
Furthermore, it is important to mention that the maximum amplitude for \mbox{CdTe-S} does not coincide with the arrival of the charge carriers at the electrode. The implications of this effect will be further discussed in sections \ref{sec:ElField} and \ref{sec:Pol}.
The observation of hole-induced current pulses in CdTe is difficult due to the very low hole mobility ($\mu_h \simeq$ 100 cm$^2$/Vs). Nevertheless, Fig. \ref{fig:Fig5.eps} shows hole signals that were recorded with a CdTe-O sensor at voltages close to the maximum bias of about 300~V and with currents below 1~$\mu$A.
 
\begin{figure}[H]
\begin{center}
\includegraphics[width=0.75\textwidth]{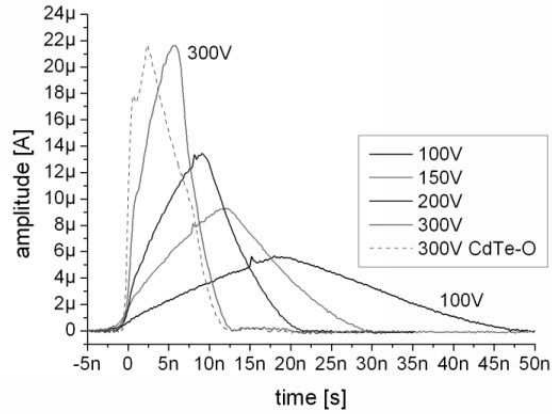}
\caption{Electron signals in \mbox{CdTe-S} with cathode irradiation. The dashed line shows a signal from an ohmic CdTe crystal (D~=~500~$\mu$m).}
\label{fig:Fig6.eps}
\end{center}
\end{figure}

\subsection{Cadmium-Zinc-Telluride}
\begin{figure}[h]
\begin{center}
\includegraphics[width=0.75\textwidth]{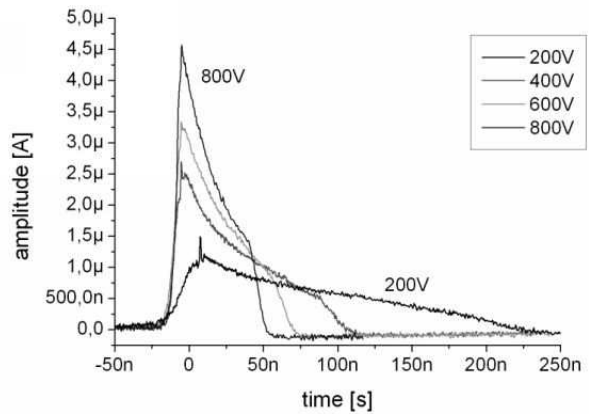}
\caption{Electron signals in CZT with cathode irradiation (D~=~2~mm).}
\label{fig:Fig7.eps}
\end{center}
\end{figure}
Figure \ref{fig:Fig7.eps} shows electron signals in a 2~mm thick CZT sample manufactured by eV-Products. The displayed curves resemble the electron signals in Si and \mbox{CdTe-O}, indicating a similar space-charge distribution in all three detector types. The current amplitude is lower than in CdTe because of the lower electric field, which also explains the lack of hole signals for CZT. The range of the applicable voltages extends up to 900~V, as above this bias the leakage current slowly increases.

\section{Collected charge}
The total collected charge is determined by a numerical integration over the recorded current pulses, which can be converted into the deposited energy. The overall precision of this energy measurement is given by three factors:
\begin{equation}
\label{eqn:Eqn2}
\sigma_{tot} = \left( \sigma_{dist}^2 + \sigma_{air}^2 + \sigma_{int}^2 \right)^{\frac{1}{2}} = \pm108~keV
\end{equation}
In the current setup the precision of the charge measurement is dominated by the error in the detector-source distance, caused by the alignment of the $^{241}$Am-source and the sensor crystal. A future improvement of the alignment precision will reduce the uncertainty in the deposited energy, but currently the error due to the detector-source misalignment is $\sigma_{dist}$ $\simeq$ 100~keV. Another factor that influences the experimental precision is the statistical fluctuation of the $\alpha$-particles energy loss in air $\sigma_{air}$, which is approximately $\pm$30~keV in the setup. The last contribution to the overall error is $\sigma_{int}$. $\sigma_{int}$ parameterizes the error in the integration over the current pulses due to variations in the length of the integration interval and is estimated to be $\pm$27~keV. In total this results in a precision of the charge collection measurements of approximately 3~$\%$.
\begin{figure}[h]
\begin{center}
\includegraphics[width=0.75\textwidth]{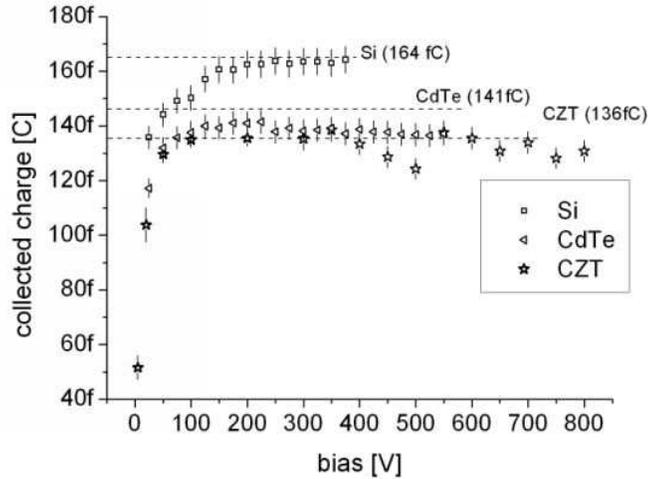}
\caption{Collected charge as a function of applied bias for different materials. Expected amount of collected charge, i.e. 100$~\%$ collection efficiency, given by dashed lines.}
\label{fig:Fig8.eps}
\end{center}
\end{figure}
The measured collected charge or energy can however be smaller than the original $\alpha$-particle energy, as charge trapping may occur.\\
As can be seen from Fig. \ref{fig:Fig8.eps}, the measured collected charges in all investigated materials saturate at higher bias. These saturation values have to be compared with the expected amount of charge that is deposited by a 3.9~MeV $\alpha$-particle (141~fC in CdTe and 136~fC in CZT). For the investigated Si p-n diodes the average electron-hole pair creation energy has been measured in \cite{Pernegger04:Lit} to be 3.8~eV~$\pm$~0.6~eV, giving a deposited charge of about 164~fC. Although this value is slightly above the established 3.62~eV, the saturation value of 164~fC in Fig. \ref{fig:Fig8.eps} supports this result. For CdTe and CZT the deposited charges agree well with the expected amount of charge, leading to the conclusion that both semiconductor materials do not show measurable electron trapping. Based on this finding, the shape of the current pulses has to be interpreted in terms of space charges (see section \ref{sec:ElField}).
From the lack of trapping also follows, that the average lifetime $\tau_{cc}$ and the average mean free path $\lambda_{cc}$ exceed the pulse duration and the detector width, respectively.

\section{Mobilities}
\label{sec:Mob}
\begin{figure}[ht]
\begin{center}
\includegraphics[width=0.75\textwidth]{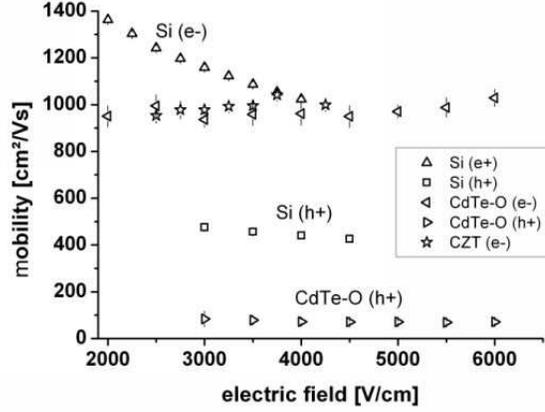}
\caption{Charge carrier mobilities in different semiconductors}
\label{fig:Fig9.eps}
\end{center}
\end{figure}

Apart from the measurement of the collected charge, the recorded current pulses also allow the determination of the charge carrier mobility $\mu$ via the pulse duration t$_E$. At this point the mathematical treatment for Si and CdTe slightly differs, as the electric field dependence of $\mu$ has to be taken into account. It has been shown, that $\mu$ deviates from its constant behavior above 2~kV/cm in Si \cite{Jacoboni76:Lit} and above approx. 12~kV/cm in CdTe \cite{Zanio71:Lit}. The condition for a non-constant mobility was only met for Si, so that two different formulae were used:
\begin{eqnarray}
\label{eqn:Eqn3}
&\mu_{CdTe/CZT} &= \frac{D^2}{t_E V}\\
&\mu_{Si} &= \frac{D^2}{2t_E V_{FD}} \cdot ln \left[ \frac{V + V_{FD}}{V - V_{FD}} \cdot \left( 1 - \frac{x_0}{D} \frac{2V_{FD}}{V + V_{FD}} \right) \right]
\end{eqnarray}
The error in the determination of the charge carrier mobility comes from the determination of the transit time t$_E$. From Fig. \ref{fig:Fig9.eps} it is evident that the charge carrier mobilities in Si show the expected field dependence and that the electron mobilities in CdTe and CZT remain constant up to 6~kV/cm. Averages over the measured mobilities yield electron mobilities of (956~$\pm$~29) cm$^2$/Vs for \mbox{CdTe-O-1} and (1022~$\pm$~24)~cm$^2$/Vs for \mbox{CdTe-O-2}, a hole mobility of (72~$\pm$~11) cm$^2$/Vs in \mbox{CdTe-O-1} and an electron mobility of (990~$\pm$~25)~cm$^2$/Vs in CZT. The values for Si lie within 5$\%$ of the values calculated by an empirical formula \cite{Jacoboni76:Lit} (at T = 27$^\circ$C). 
For CZT the measured electron mobilities agree well with the 1000~cm$^2$/Vs stated by the manufacturer. Similar details on the electron and hole mobilities in the CdTe samples were not available from the manufacturer, so that the measured values have to be compared with the results of other groups. These literature values \cite{Zanio71:Lit,Turkevych02:Lit} lie around $\mu_e$ = (1000~-~1100) cm$^2$/Vs and $\mu_h$ $\simeq$ 100 cm$^2$/Vs, well in agreement with the results reported here.

\section{Electric Field}
\label{sec:ElField}
The electric field profile in a single channel sensor crystal can be deduced from a current pulse under the premise of a point-like charge cloud and the absence of charge carrier trapping. With this prerequisite it is possible to calculate the charge carrier position x(t) from the numerical integral over a current pulse:
\begin{equation}
\label{eqn:Eqn4}
Q(t_E) = \int^{t_E}_0 i(t)dt = \frac{Q_0^*}{D} \cdot \int^{t_E}_0 \frac{dx(t)}{dt} dt = \frac{Q_0^*}{D} \cdot \left( x(t_E) - x(0) \right) 
\end{equation}
Q$_0^*$ is the integrated charge up to the time t$_E$, where the bend in the current signal occurs and Q$_0$ is the total collected charge. From (\ref{eqn:Eqn4}) follows the electric field profile along the charge carriers path:
\begin{equation}
\label{eqn:Eqn5}
E(x(t)) = \frac{i(x(t)) \cdot D}{Q_0 \cdot \mu}
\end{equation}  

\subsection{Si}
\label{subsec:SiFeld}
\begin{figure}[ht]
\begin{center}
\includegraphics[width=0.75\textwidth]{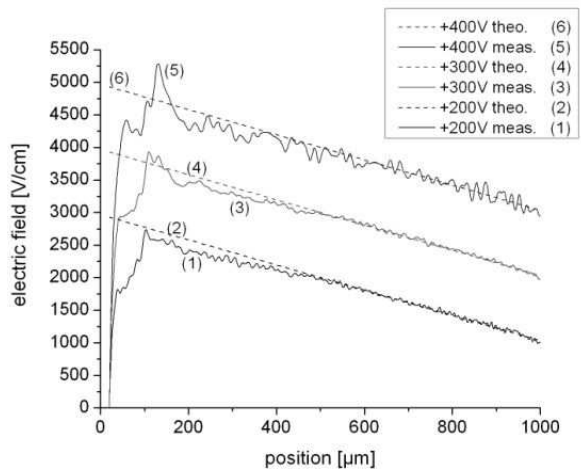}
\caption{Measured (solid curves) and theoretical (dashed lines) electric field distributions in the n-type region of a 1~mm thick Si p+ n diode for three different voltages. Electrons start at the left and progress toward the anode (right).}
\label{fig:Fig10.eps}
\end{center}
\end{figure}

Fig. \ref{fig:Fig10.eps} shows the measured (solid lines) and the theoretical (dashed lines) electric field profiles in the n-doped part of the Si p+ n diodes at three different bias settings. The theoretical field distributions \cite{Lutz99:Lit} were calculated with a depletion voltage \mbox{V$_{FD}$ = 96~V} and a starting position of the charge carriers \mbox{x(0) = 20~$\mu$m}. As expected, the measured field strengths decrease linearly from the cathode towards the anode, based on the constant positive space-charge density in the fully depleted n-type material. Fits to the constant slopes of the field profiles give a space-charge density of (3.3~$\pm$~0.65)~x~10$^{10}$~cm$^{-3}$.
The theory of the p-n junction does not support the presence of a negative space-charge inside the n-doped material. Therefore the deviations of the measured electric field from the theoretical expectation close to the cathode have to be explained by two other effects.
\begin{itemize}
\item{i) }Limited bandwidth of the TCT-setup. The bandwidth of the system has been tested by injecting voltage pulses with rise times of 500~ps into the circuit. The systems overall bandwidth is measured to be about 500~MHz. Therefore current pulse rise times of more than 3~ns cannot be explained by the electronics alone.
\item{ii) }Plasma effect. One $\alpha$-particle creates approx. 1.2 million electron-hole pairs along a cylindrical track of about 10-20~$\mu$m length. These charge carriers are not separated instantly as they shield the external electric field, which then causes the charge carrier migration to start with a delay on the order of several hundred picoseconds to several nanoseconds. Different authors \cite{Kanno89:Lit,Galster85:Lit,Seibt73:Lit,Alberigi68:Lit} measured the influence of the plasma effect on the charge collection for $\alpha$-particles with energies below 10~MeV.
\end{itemize}
Therefore the distortion of the measured electric field close to the cathode has to be a feature of the experimental method and not of the material itself. This experimental limitation of the transient current technique is hard to overcome, as TCT measurements need large signal charges. This means that only $\alpha$-particles and lasers are valid signal sources, but the use of a laser is problematic because of the surface treatment and the metal electrodes of the crystals. Nevertheless, the very good agreement between prediction and measurement shows, that the determination of the electric field from a TCT-measurement is valid.

\subsection{CdTe-O}
\label{subsec:E-CdTeO}
\begin{figure}[h]
\begin{center}
\includegraphics[width=0.75\textwidth]{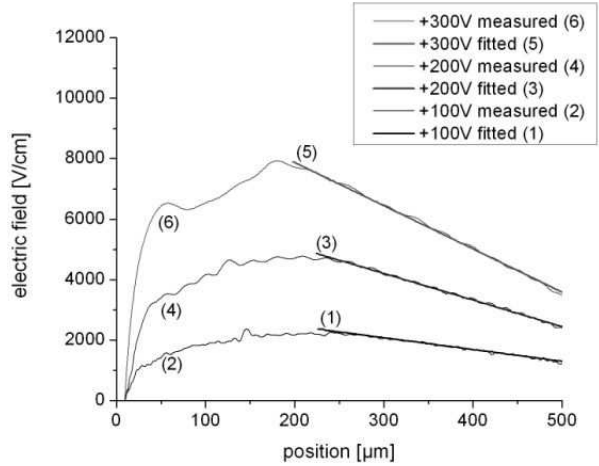}
\caption{Measured and fitted electric field profiles inside a 500~$\mu$m thick \mbox{CdTe-O} sensor. Anode on the right.}
\label{fig:Fig11.eps}
\end{center}
\end{figure}

Figure \ref{fig:Fig11.eps} shows the measured electric field profiles in \mbox{CdTe-O} at three different voltages. All curves display a maximum close to the middle of the detector, followed by a linear decrease in field strength towards the anode. From Fig.~\ref{fig:Fig11.eps} can be seen, that the electric field profile to the right of the maximum follows a linear behavior. Using Poisson's equation the space-charge density in this region can be calculated. The results are given in Table \ref{Table}.\\
However, the electric field deviates from this linear behavior close to the cathode. This field profile can be explained by two effects. First, the presence of two oppositely charged space charges inside the detector. That is, a negative space-charge in front of the cathode and a positive one next to the anode. The origin of these opposing charges could be the injection of charge carriers through both ohmic contacts \cite{Mayer65:Lit}. Second, the plasma effect, which is the main reason for the deviation of the electric field profile close to the cathode in Si.\\
In order to determine which effect dominates the electric field behavior it was tried to reconstruct the measured current pulses under the assumption that only the plasma effect is responsible for the deviations close to the cathode and that the true electric field inside the crystals is a linear extrapolation of the fits in Fig.~\ref{fig:Fig11.eps}. The reconstructed current pulses were calculated based on the Ramo-Shockley theorem and a simple, first-order model of the plasma effect, based on two time-constants:
\begin{itemize}
\item{i) }Reduced signal charge and charge carrier velocity. As the dense charge cloud erodes the shielding effect is weakened. Thus, the number of charge carriers Q(t) that contribute to the current signal rises exponentially with time. Additionally the reduced electric field inside the charge cloud causes a reduction in the charge carrier velocity. Therefore, the charge carriers do not move with the velocity dominated by the externally applied bias, but rather pick up speed as the density of the charge cloud degrades. These two effects are parameterized by exponential functions with the time-constant $\tau$. A much more detailed model can be found in \cite{Kanno89:Lit}.
\item{ii) }Delayed signal formation. The movement of the charge carriers does not start immediately after their generation, but is delayed by several nanoseconds \cite{Seibt73:Lit,Alberigi68:Lit}. This shift is given by t$_{s}$.
\end{itemize}
\begin{eqnarray}
&&i(t) = Q(t) \cdot \frac{\mu}{D} \cdot E(x(t))
 = Q_0 \cdot ( 1 - e^{\frac{t}{\tau}}) \cdot  \frac{\mu}{D} \cdot ( a \cdot 
 x(t) + b ) \\
&&E(x(t)) = a \cdot x(t) + b\\
&&x(t) = \left( \frac{b}{a} + (x_0 - \frac{b}{a}) \cdot e^{-a \cdot \mu \cdot (t - t_s)} \right) \cdot ( 1 - e^{\frac{t}{\tau}})
\end{eqnarray}
\begin{figure}[ht]
\begin{center}
\includegraphics[width=0.75\textwidth]{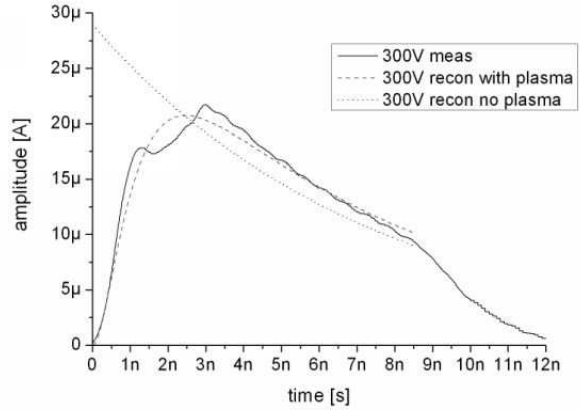}
\caption{Measured and reconstructed current pulses in a 500~$\mu$m thick \mbox{CdTe-O} sensor at 300~V bias.}
\label{fig:Fig12.eps}
\end{center}
\end{figure}\\
Fig.~\ref{fig:Fig12.eps} shows the results of the reconstruction at 300~V bias. The solid line represents the measured curve, whereas the dotted line indicates the expected current signal without the above model for the plasma effect. The dashed line shows the fully reconstructed signals with plasma effect. The observed time-constants are comparable to the values reported by \cite{Seibt73:Lit,Alberigi68:Lit}. In addition, the measurements support an inverse electric field dependence of the plasma effect, which was also found by \cite{Seibt73:Lit,Alberigi68:Lit}. Although a final answer regarding the electric field profile close to the cathode can not be given, the results of the current pulse reconstruction indicate, that the plasma effect can not be discarded as a reason for the observed field behavior.\\
As a conclusion, the agreement between the electric field profiles presented herein and the results from other electric field measurements \cite{Manfredotti96:Lit,Hage-Ali94:Lit} proves, that the electric field strength in CdTe-O has a maximum in the vicinity of the cathode.

\begin{table*}[h]
\begin{center}
\begin{tabular}{|l|l|c|c|c|c|c|c|}
\hline
    CdTe-O &     V$_{bias}$ [V] &        100 &        150 &        200 &        250 &        300 &            \\
\hline
           & N$_D$ [$10^{11}$~cm$^{-3}$] &   2.37  &   4.07 &   5.35 &   7.35 &   8.69 &            \\
\hline
           & $\Delta$N$_D$ [$10^{9}$~cm$^{-3}$] &   1.88  &   2.35 &   2.39 &   1.34 &   2.37&            \\
\hline
           &   t$_s$ [ns] &        4.0 &        2.3 &        1.6 &        1.0 &        0.85 &            \\
\hline
           &   $\tau$ [ns] &        4.0 &        2.2 &        1.6 &        0.9 &        0.8 &            \\
\hline
       CZT &     V$_{bias}$ [V] &        300 &        400 &        500 &        600 &        700 &        800\\
\hline
           & N$_D$ [$10^{11}$~cm$^{-3}$] &   0.61 &   0.97 &   1.22 &   1.28 &   1.46 &   1.58\\
\hline
           & $\Delta$N$_D$ [$10^{8}$~cm$^{-3}$] &   3.04 &   4.86  &   6.56 &   3.83 &   3.61 &   4.09      \\
\hline
\end{tabular}  
\end{center}
\caption{Space-charge densities in \mbox{CdTe-O} and CZT. The model parameters t$_s$ and $\tau$ were obtained from fits with $\delta$t$_s$~$\simeq$~0.2~ns and $\delta$t$_s$~$\simeq$~0.1~ns.}
\label{Table}
\end{table*}

\subsection{CdTe-S}

As stated above, the calculation of the electric field relies on the determination of the arrival time t$_E$. For \mbox{CdTe-S} this is more complicated than for \mbox{CdTe-O} and CZT, because the \mbox{CdTe-S} crystals do not show an easily identifiable arrival of the electrons at the anode. Although the curves show a pronounced maximum, it is unlikely that this bend in the current signals indicates the arrival time t$_E$.\\
The basis for this assumption is that the two crystal types differ only in their contact electrodes. From (\ref{eqn:Eqn1}) then follows, that a difference in the current amplitude can only be caused by a difference in the electric field profile. The direct comparison of the current pulses in Fig. \ref{fig:Fig6.eps} shows, that the maximum and average amplitudes in \mbox{CdTe-O} and \mbox{CdTe-S} are equal, which also means, that the average transit times have to be comparable. Therefore the average signal durations (transit time t$_E$) in \mbox{CdTe-S} were taken to be in the order of the respective signal durations in \mbox{CdTe-O}.
\begin{figure}[ht]
\begin{center}
\includegraphics[width=0.75\textwidth]{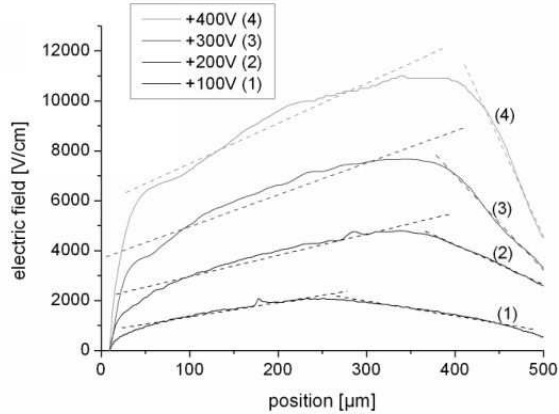}
\caption{Measured (solid lines) and fitted (dashed lines) electric field profiles inside a 500~$\mu$m thick \mbox{CdTe-S} sensor. Anode on the right.}
\label{fig:Fig13.eps}
\end{center}
\end{figure}
Based on this finding Fig. \ref{fig:Fig13.eps} shows the electric field profiles in \mbox{CdTe-S} for different bias settings. In contrast to the field profile in \mbox{CdTe-O} the electric field in \mbox{CdTe-S} has a minimum at the cathode from where it rises, before it falls off again close to the anode. The rising portion of the electric field agrees with the expectation, as the Schottky-contact of these sensors is good enough to be blocking for holes \cite{Tak99:Lit,Tak01:Lit}. Consequently the ohmic cathode injects electrons into the crystal, generating a negative space-charge and a rising electric field. The measured break-down of the electric field close to the anode does not agree with this expectation, but can again be interpreted in terms of a positive space-charge. The reason for this positive space-charge might be the enhanced emission of holes through the anode into the crystal at high electric fields. In that case the conduction properties of the sensor would be space-charge and recombination controlled \cite{Kao84:Lit}.

\subsection{CZT}
\begin{figure}[h]
\begin{center}
\includegraphics[width=0.75\textwidth]{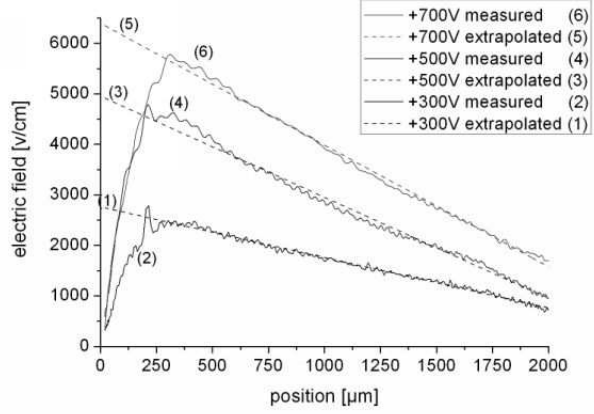}
\caption{Measured (solid lines) and extrapolated (dashed lines) electric field profiles inside a 1~mm thick CZT sensor. Anode on the right.}
\label{fig:Fig14.eps}
\end{center}
\end{figure}
Figure \ref{fig:Fig14.eps} shows the electric field distribution inside a 2~mm thick CZT crystal. The situation is similar to \mbox{CdTe-O} with the maximum field strength close to the cathode and a linearly decreasing electric field towards the anode.

\section{Polarization}
\label{sec:Pol}

\begin{figure}[H]
\begin{center}
\includegraphics[width=0.75\textwidth]{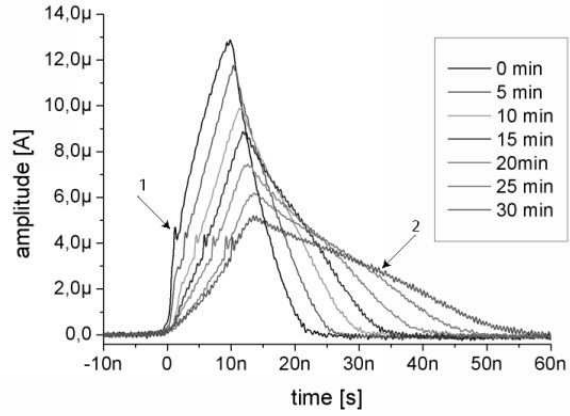}
\caption{Effect of polarization on current signals at 200~V in a 500~$\mu$m thick \mbox{CdTe-S} sensor. Over 30 minutes the signal amplitude decreases significantly, whereas the signal duration increases.}
\label{fig:Fig15.eps}
\end{center}
\end{figure}

\begin{figure}[H]
\begin{center}        
\includegraphics[width=0.75\textwidth]{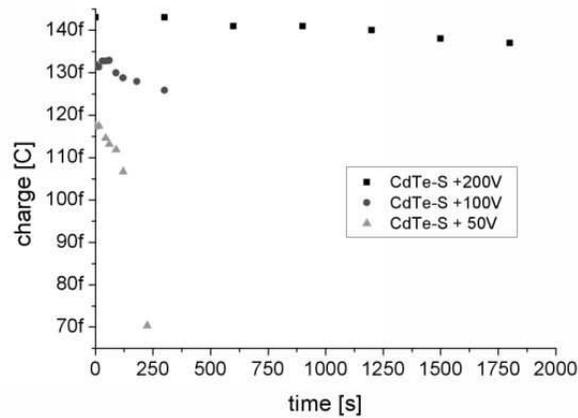}
\caption{Measured collected charge over time for three different bias settings. The measurements were performed with a charge sensitive setup and a 500~$\mu$m thick \mbox{CdTe-S} sensor.}
\label{fig:Fig16.eps}
\end{center}
\end{figure}

\begin{figure}[H]
\begin{center}
\includegraphics[width=0.75\textwidth]{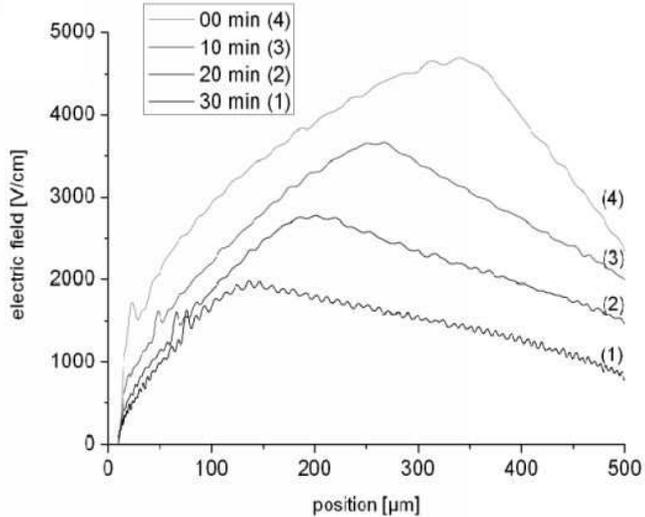}
\caption{Changes in the electric field of a 500~$\mu$m thick \mbox{CdTe-S} sensor due to polarization at +200~V. Anode on the right.}
\label{fig:Fig17.eps}
\end{center}
\end{figure}

During the measurements it became evident, that the \mbox{CdTe-S} sensors suffer from a significant decrease in signal amplitude, if they are operated at a low voltages for a longer time. This can be attributed to the polarization of the detector, i.e. the gradual accumulation of fixed space-charges inside the detector. These space-charges deform the electrical field up to a point where the operation of the detector becomes impossible. None of the other tested materials showed this behavior, which is why this section will only deal with the \mbox{CdTe-S} sensors.\\
Figure \ref{fig:Fig15.eps} shows the effect of the polarization on the current pulses. With ongoing operation of the detector the initial knee (1.st arrow) in the pulse shape is lost and the current amplitude is reduced. In addition to this, the position of the maximum current amplitude with respect to the beginning of the signal is barely changed. This means, that although the overall electric field strength in the sensor is reduced, the maximum field strength is still reached after the same time. Considering the reduced charge carrier velocity under a reduced field strength, it follows that the position of the maximum electric field strength in the sensor is shifted towards the cathode. If the position of the maximum was fixed, i.e. would lie at the anode, it should take more than twice as long for it to be reached after 30 mins of operation as compared to 0 mins. This is clearly not supported by the observed current pulses.\\
The second feature of the current signals is that they develop a bend in the falling slope (2.nd arrow). This and the change in the position of the maximum electric field strength both speak in favor of the argument, that the charge carriers reach the electrode only after the maximum current amplitude has occurred.\\
The effect of the polarization is also visible in the collected charge (see Fig. \ref{fig:Fig16.eps}). At 50~V bias the charge-sensitive setup stopped recording the pulses after only 4~mins of operation. Switching off the bias for a few seconds neutralizes the polarization and causes the process to begin again. 
\newpage
Using the same formalism as in section \ref{sec:ElField} it is possible to observe the time-wise behavior of the electric field inside a \mbox{CdTe-S} sensor (see Fig. \ref{fig:Fig17.eps}). The figure shows that the externally applied electric field is gradually compensated and that the position of the maximum field strength is shifted from the middle of the sensor crystal towards the cathode. A possible explanation for the shift of the maximums position could again be the injection of holes through the anode and the time-dependent expansion of this space-charge towards the cathode. These results are in contrast to the models \cite{Bell74:Lit,Siffert76:Lit}, which assume the build-up of a negative space-charge in front of the anode and the resulting reduction of the electric field. The models further imply that the major part of the detector volume shows a low field strength and that only the region in front of the anode possesses a high electric field. 

\section{Conclusions}

In summary, this paper presented a TCT-study of Si, CdTe and CZT. It was found, that the investigated crystals do not show measurable electron trapping, indicating a material quality that is suitable for particle detection. Furthermore it was shown that the electric field profiles in ohmic CdTe, in Schottky-contacted \mbox{CdTe-S} and in CZT can be approximated by linear relations and that all detectors feature space-charges inside the bulk material. Finally, it was demonstrated that Schottky-contacted CdTe sensors show a significant polarization at low voltages, which makes the operation at high bias necessary.

\section{Acknowledgements}
The authors gratefully acknowledge the provision of material samples by T. Takahashi (University of Tokyo).

\bibliographystyle{unsrt}
\bibliography{Lit_eng}

\begin{thebibliography}{10}

\bibitem{Ramo39:Lit}
S.~Ramo.
\newblock {"Currents induced by electron motion"}.
\newblock {\em Proc. of the I.R.E.}, 27:584, 1939.

\bibitem{Shockley38:Lit}
W.~Shockley.
\newblock {"Currents to conductors induced by a moving point charge"}.
\newblock {\em J. Appl. Phys.}, 9:635, 1938.

\bibitem{Lutz99:Lit}
G.~Lutz.
\newblock {\em {"Semiconductor radiation detectors"}}.
\newblock {Springer}, 1999.

\bibitem{Zanio68:Lit}
K.~R. Zanio et~al.
\newblock {"Transient Currents in Semi-Insulating CdTe Characteristic of Deep
  Traps"}.
\newblock {\em J. Appl. Phys.}, 39(6):2818 -- 2828, 1968.

\bibitem{Krs04:Lit}
Olaf Krasel.
\newblock {\em {"Charge collection in irradiated silicon-detectors"}}.
\newblock PhD thesis, {University of Dortmund}, Juli 2004.

\bibitem{Jacoboni76:Lit}
C.~Jacoboni.
\newblock {"A review of some charge transport properties of silicon"}.
\newblock {\em Solid State Electronics}, 20:77 -- 89, 1976.

\bibitem{Pernegger04:Lit}
H.~Pernegger et~al.
\newblock {"Charge-carrier properties in synthetic single-crystal diamond
  measured with the transient-current technique"}.
\newblock {\em J. Appl. Phys.}, 97, 2005.

\bibitem{Zanio71:Lit}
K.~R. Zanio et~al.
\newblock {"Transport properties in CdTe"}.
\newblock {\em Phys. Rev. B}, 4(2):422 -- 431, 1971.

\bibitem{Turkevych02:Lit}
I.~Turkevych et~al.
\newblock {"High-temperature electron and hole mobility in CdTe"}.
\newblock {\em Semicond. Sci. Technol.}, 17:1064 -- 1066, 2002.

\bibitem{Kanno89:Lit}
I.~Kanno.
\newblock {"A model of charge collection in a silicon surface barrier
  detector"}.
\newblock {\em Rev. Sci. Instrum.}, 61:129 -- 137, 1989.

\bibitem{Galster85:Lit}
W.~Galster et~al.
\newblock {"The influence of plasma effects on the timing properties of
  surface-barrier detectors for heavy ions"}.
\newblock {\em Nucl. Instrum. Methods A}, 240:145 -- 151, 1985.

\bibitem{Seibt73:Lit}
W.~Seibt et~al.
\newblock {"Charge collection in silicon detectors for strongly ionizing
  particles"}.
\newblock {\em Nucl. Instrum. Methods}, 113:317 -- 324, 1973.

\bibitem{Alberigi68:Lit}
A.~Alberigi~Quaranta et~al.
\newblock {"Plasma time and related delay effects in solid state detectors"}.
\newblock {\em Nucl. Instrum. Methods}, 72:72 -- 76, 1968.

\bibitem{Mayer65:Lit}
J.~W. Mayer et~al.
\newblock {"Observation of Double Injection in Long Silicon p-i-n Structures"}.
\newblock {\em Phys. Rev.}, 137:286 -- 294, 1965.

\bibitem{Manfredotti96:Lit}
C.~Manfredotti et~al.
\newblock {"Investigation on the electric field profile in CdTe by ion beam
  induced current"}.
\newblock {\em Nucl. Instrum. Methods A}, 380:136 -- 140, 1996.

\bibitem{Hage-Ali94:Lit}
M.~Hage-Ali et~al.
\newblock {"Internal field distribution in CdTe detectors prepared from
  semi-insulating materials"}.
\newblock {\em Proc. SPIE}, 2305:157 -- 161, 1994.

\bibitem{Tak99:Lit}
Tadayuki Takahashi et~al.
\newblock {"High resolution Schottky CdTe diode for hard x-ray and gamma-ray
  astronomy"}.
\newblock {\em Nucl. Instrum. Methods A}, 436:111 -- 119, 1999.

\bibitem{Tak01:Lit}
Tadayuki Takahashi et~al.
\newblock {"High resolution Schottky CdTe diode detector"}.
\newblock {\em IEEE Trans. Nucl. Sci.}, 49(3):1297 -- 1303, 2002.

\bibitem{Kao84:Lit}
K.~C. Kao.
\newblock {"Double injection in solids with non-ohmic contacts I: Solids
  without defects"}.
\newblock {\em J. Phys. D}, 17:1433 -- 1448, 1984.

\bibitem{Bell74:Lit}
R.~O. Bell et~al.
\newblock {"Time-Dependent Polarization of CdTe Gamma-Ray Detectors"}.
\newblock {\em Nucl. Instrum. Methods}, 117:267--271, 1974.

\bibitem{Siffert76:Lit}
P.~Siffert and R.~O. Bell.
\newblock {"Polarization in cadmium telluride nuclear radiation detectors"}.
\newblock {\em IEEE Trans. Nucl. Sci.}, 23(1):159 -- 170, 1976.

\end{thebibliography}

\end{document}